\documentclass[prl,twocolumn,preprintnumbers,amsmath,amssymb]{revtex4}

\usepackage{bm,curves}
\usepackage[colorlinks=true,linkcolor=blue]{hyperref} 
\usepackage{color}
\usepackage{dcolumn}
\usepackage{ulem}
\def\xIC{\langle x \rangle_{_{\rm IC}}}

\begin{document}

\preprint{JLAB-THY-15-2031}

\title{Reply to Comment on ``New limits on intrinsic charm in the
	nucleon	from \\ global analysis of parton distributions''}

\author{P. Jimenez-Delgado$^1$,
        T. J. Hobbs$^2$,
        J. T. Londergan$^3$,
        W. Melnitchouk$^1$}

\affiliation{
	$^1$\mbox{Jefferson Lab, 12000 Jefferson Avenue,
         Newport News, Virginia 23606, USA} \\
        $^2$\mbox{Department of Physics,
         University of Washington, Seattle, Washington 98195, USA} \\
        $^2$Department of Physics and
         Center for Exploration of Energy and Matter,
         Indiana University, Bloomington, Indiana 47405, USA}

\date{\today}

\begin{abstract}
We reply to the Comment of Brodsky and Gardner on our paper 
``New limits on intrinsic charm in the nucleon from global analysis
of parton distributions'' [Phys. Rev. Lett. {\bf 114}, 082002 (2015)].
We address a number of incorrect claims made about our fitting
methodology, and elaborate how global QCD analysis of all available
high-energy data provides no evidence for a large intrinsic charm
component of the nucleon.
\end{abstract}
\maketitle

%%%%%%%%%%%%%%%%%%%%%%%%%%%%%%%%%%%%%%%%%%%%%%%%%%%%%%%%%%%%%%%%%%%%%%%%
In a recent Comment \cite{BG}, Brodsky and Gardner (BG) criticize
our global PDF analysis \cite{JDHLM} of all available high-energy
scattering data, including those from fixed-target experiments
at high $x$ and low $Q^2$, which placed strong constraints
on the magnitude of intrinsic charm (IC) in the nucleon.
For a range of models of IC, the analysis \cite{JDHLM} strongly
disfavored large magnitudes of IC, with the momentum fraction
carried by charm quarks $\xIC$ at most 0.5\% at the 4$\sigma$
confidence level (CL).

% ......................................................................
% \Delta\chi^2

BG claim that because our global analysis \cite{JDHLM} uses
${\cal O}(30)$ parameters, as is typical in all such fits, one must
adopt a much larger tolerance criterion than $\Delta\chi^2 = 1$.
In fact, it is well known that for Gaussian distributions parameter
errors in $\chi^2$ fits are determined by $\Delta\chi^2 = 1$,
irrespective of the number of parameters in the fit \cite{PDG, NumRec}.
The parameter $m$ in Table~38.2 of Ref.~\cite{PDG}, for example,
% cited by BG in v1
is the dimensionality of the error regions for joint distributions
($m=1$ for linear errors, $m=2$ for error ellipses, {\it etc.}),
and has nothing to do with the total number of parameters in the fit.
Actually, Fig.~38.2 of Ref.~\cite{PDG} involves the number of degrees
of freedom of a fit (number of points $-$ number of parameters)
and not the number of parameters in the fit.
For the determination of individual parameter errors, the correct
dimension is $m=1$, which gives $\Delta\chi^2=1$ at the 68.3\% CL.
(For examples of error ellipses with $m=2$, see Fig.~12 of
Ref.~\cite{JR}.)

The parameter errors and $\chi^2$ profiles related to one-dimensional
probablility distributions are correctly evaluated using
$\Delta\chi^2 = 1$.  Errors on other quantities are then computed
using standard error propagation techniques, such as the Hessian
method; they can also be used to produce error regions of different
dimensionalities with the appropriate $\Delta\chi^2$ criteria
\cite{PDG, NumRec}.
Apparently, BG have confused the dimensionality of error regions with
the number of independent parameters in a fit.  Their claims about
$\Delta\chi^2$ are simply wrong.

Tolerance criteria $\Delta\chi^2 > 1$ are used by some PDF groups
\cite{MSTW, CJ, CT} on purely phenomenological grounds, to account
for tensions among different data sets, while others
\cite{JR, HERAPDF, ABM} use the standard $\Delta\chi^2 = 1$.
The $\chi^2$ profiles in \cite{JDHLM} were presented as a function
of $\xIC$, so that $\xIC$ values for different tolerance choices
can be easily compared.
BG also suggest that our single parameter errors were obtained
by fixing the other parameters at the $\chi^2$ minimum.
This is not true: we minimize the $\chi^2$ with respect to all
other parameters in the fit, as is standard procedure in global fits.
Had we not properly refitted the complete model, the rise in the
$\chi^2$ away from the minimum would be even steeper than for the
profiles shown in Fig.~1 of Ref.~\cite{JDHLM}.

% ......................................................................
% SLAC data

Inclusive DIS cross sections, such as those measured at SLAC, receive
contributions from all quark flavors, so they cannot by themselves
provide significant constraints on charm.  The power of a global fit,
however, lies in the correlation between different observables,
with different weightings of quark flavors, within the framework
of perturbative QCD.
While the bulk of the data from SLAC \cite{SLAC} at large $x$ lie
below the charm threshold, cross sections below threshold constrain
light quark distributions, which indirectly impacts on the
determination of IC at the same kinematics.
Our analysis also accounts for the suppression of charm
production below and near the hadronic charm threshold
\cite{JDHLM, BG}.
Implementing the suppression involves some model dependence
in relating the partonic and hadronic charm thresholds
\cite{JDHLM, MSTW}, and while this affects the quantitative limits
(with partonic threshold factors alone $\xIC$ would be $< 0.1\%$
at the 5$\sigma$ CL), the effects do not alter the overall
conclusions about the magnitude of IC supported by the data.

To avoid dealing with complications from thresholds and other
hadronic effects at low $W^2$ and $Q^2$, many global PDF analyses
impose more severe cuts on $W^2$ and $Q^2$ than those in
Ref.~\cite{JDHLM}.  While this simplifies the theoretical
treatment, it also removes a significant amount of data at
large $x$ that could potentially impact the question of IC.

Recently, some PDF analyses \cite{JR, ABM, CJ} have relaxed the
conventionally more restrictive $W^2$ and $Q^2$ cuts in order
to better constrain large-$x$ PDFs.
Such analyses benefit from increased statistics at large $x$,
but require careful treatment of subleading $1/Q^2$ and nuclear
corrections.  Our analysis \cite{JDHLM} employs the standard
treatment of target mass corrections (TMCs) \cite{SBMS12},
phenomenological higher twists determined consistently within
the same fit \cite{JR}, and the latest technology in nuclear
corrections \cite{CJ}.
Apparently confusing Refs.~\cite{JR} and \cite{CJ}, BG assert that
we model higher twists as isospin independent, and that our TMCs
are problematic at $x\to 1$.  In fact, our higher twist corrections
do depend on isospin, as evident from Table~III of Ref.~\cite{JR},
and are determined empirically without assuming any functional form.

Furthermore, the well-known threshold problem of TMCs at $x=1$
is relevant only at very low $W^2$, well below the cuts made in
all global PDF analyses \cite{SBMS12}.  It is also not true that
we neglect intrinsic strangeness and bottom: the $s$ and $\bar s$
PDFs are parametrized model independently at the input scale,
and given our results for IC, intrinsic bottom is negligible
for the current phenomenology \cite{Lyonnet15}.

Our global fit carefully propagates all statistical and
systematic errors, both uncorrelated and correlated,
including normalization, for all data sets used.
(For details of the fitting code see Ref.~\cite{Chromopolis}.)
For the SLAC data, our analysis uses the original hydrogen and
deuterium cross sections \cite{SLAC} rather than the derived
structure functions (obtained by combining measurements at
different energies), which allows for a more exact treatment
of point to point correlated errors.
Aside from the SLAC data, other measurements, such as the NMC
proton and deuteron cross sections \cite{NMC_sig} and the
inclusive proton cross sections from HERA \cite{HERA_sig},
also disfavor nonzero values of IC.

% ......................................................................
% EMC data

In addition to the fit of the standard high energy data sets used
by most PDF groups, in Ref.~\cite{JDHLM} we also considered a fit
including data from the EMC measurement of the charm structure
function $F_2^c$ \cite{EMC_F2c} --- sometimes cited \cite{Brodsky}
as providing evidence for large IC in the nucleon.  In practice,
the EMC data have strong tension with other measurements, and
give a very large overall $\chi^2/N_{\rm dat}$ of $\gtrsim 4$,
and a $Q^2$ dependence incompatible with perturbative QCD.
Several of the EMC data points at the highest $x$ values
($x \gtrsim 0.2$), where there are no other direct constraints
from charm production experiments, lie systematically above all
global fits, including ones with IC contributions \cite{JDHLM}.
At the same time, at low $x$ values ($x \lesssim 0.02$) where charm
distributions are strongly constrained by HERA \cite{HERA_F2c},
however, the EMC data are significantly below the fitted results.

We thus disagree with the assertion of BG that $\xIC \sim {\cal O}(1\%)$
is ``consistent with the analysis of the EMC measurements'' \cite{BG}.
No reasonable amount of nuclear corrections (which are, in fact,
considered in Ref.~\cite{JDHLM}) or $\Delta\chi^2$ tolerance can
reconcile the EMC $F_2^c$ data with the rest of the global data set
within a QCD framework, without invoking a very peculiar shape for
IC that is strongly at variance with all IC models \cite{JDHLM, BHPS,
Hobbs14}.  Consequently, no modern QCD analysis
\cite{JR, MSTW, CJ, CT, HERAPDF, ABM, Pumplin, Dulat, NNPDF}
includes the EMC data in their fits.
The MSTW analysis \cite{MSTW} compared $F_2^c$ computed from
their PDFs with the EMC charm measurements, and concluded that
``If the EMC data are to be believed, there is no room for a very
sizable intrinsic charm contribution.''
We agree with this conclusion.

%%%%%%%%%%%%%%%%%%%%%%%%%%%%%%%%%%%%%%%%%%%%%%%%%%%%%%%%%%%%%%%%%%%%%%%%%
This material is based upon work supported by the National Science
Foundation (T.J.H. and J.T.L.) under Grant PHY-1205019. The work of
T.J.H. was also supported in part by DOE Grant No. DE-FG02-87ER40365
and DE-FG02-97ER-41014.  P.J.-D. and W.M. were supported by the DOE
Contract No. DE-AC05-06OR23177, under which Jefferson Science
Associates, LLC operates Jefferson Lab.

%%%%%%%%%%%%%%%%%%%%%%%%%%%%%%%%%%%%%%%%%%%%%%%%%%%%%%%%%%%%%%%%%%%%%%%%%

\end{document}